\documentclass[aps,prb,twocolumn,amsmath,amssymb]{revtex4-2}

\usepackage{graphicx}
\usepackage{dcolumn}
\usepackage{bm}
\usepackage{hyperref}
\usepackage{amsmath}
\usepackage{amssymb}

\begin{document}

\title{Spatial Offset of Excited States in Non-Hermitian Lattices}

\author{Xiaohan Jiang}
\author{Yuanyuan Pan}
\author{Yang Zhang}
\author{Ye Xiong}
\thanks{Corresponding author}
\email{xiongye@njnu.edu.cn}
\affiliation{Institute of Theoretical Physics, Nanjing Normal University - Nanjing 210023, PRC}

\begin{abstract} We investigate the behavior of light-wave packets injected into
	non-Hermitian microcavity lattices under highly dissipative
	conditions. While all eigenstates of the lattice exhibit exponential
	decay, a specifically excited state maintains coherent propagation. In
	a one-dimensional lattice, this state undergoes a spatial
	displacement shift away from the injection position, which is
	a fundamental property of non-Hermitian systems with a point
	gap when the spectrum encircles a finite region in the
	complex plane. Extending such a shift to two-dimensional lattices reveals
	a geometrically anomalous V-shaped wave packet formation with orientation-tunable arms.
        Notably, this geometric control mechanism
	enables all-optical steering of non-Hermitian photonic states without
	requiring structural modifications. 

\end{abstract}

\maketitle

\section{Introductions.} 

Non-Hermitian systems governed by complex Hamiltonians exhibit unique phenomena,
including PT symmetry
breaking\cite{PhysRevLett.80.5243,feng_non-hermitian_2017,el-ganainy_non-hermitian_2018,ozdemir_paritytime_2019,doi:10.1126/science.aar7709,PhysRevA.2.1170,PhysRevLett.125.237401},
exceptional
points\cite{berry_physics_2004,Heiss_2004,Heiss_2012,PhysRevLett.117.110802,RevModPhys.93.015005,Xiong_2018},
and generalized Brillouin
zones\cite{PhysRevLett.121.086803,PhysRevB.104.165117,PhysRevLett.125.226402}.
These features fundamentally stem from the non-orthogonality between left and
right eigenstates and the eigenspectrum in the complex plane. For example,
eigenstates with positive imaginary eigenvalues (amplification) or negative
imaginary eigenvalues (dissipation) exhibit asymmetric amplification profiles in
time and space, inducing non-reciprocal transport in systems with non-reciprocal
couplings or hoppings\cite{PhysRevA.108.052205, PhysRevApplied.10.047001,
longhi_robust_2015,
PhysRevB.103.054203,PhysRevB.105.024303,PhysRevB.108.115404,PhysRevA.83.060101,Poli_2024,PhysRevB.110.245103}.
However, practical implementations face a fundamental stability dichotomy:
systems with eigenspectral components above the real axis in the complex plane
exhibit dynamical instability, while those with entirely sub-real-axis spectra
suffer from global mode decay, rendering stable experimental observation
unattainable\cite{braghini2024stability}. 

In this study, we examine the response of a non-Hermitian microcavity lattice when an external light source is injected at a specific point. To ensure stability, we apply global damping to all microcavities, which shifts the entire spectrum downward along the imaginary axis. The system's evolution is governed by a Schrödinger-like equation:
\begin{equation}
i\partial_t \vec\psi(t) = H \vec\psi(t) + \vec\psi_0(t),
\end{equation}
where $\vec\psi(t)$ is the system's state, $H$ is the non-Hermitian Hamiltonian, and $\vec\psi_0(t)$ represents the external source. Our focus is on the steady-state solution of this equation, which describes the long-lived response of the system rather than the eigenstates of $H$.

The key observation is the ``source-center offset" phenomenon. In conventional
systems, wave packets propagate symmetrically from the external source location,
akin to ripples in water spreading from where a stone is dropped. In contrast,
our one-dimensional (1D) non-Hermitian lattice exhibits a wave packet emerging
from a point offset from the physical injection position. This offset
is not a topological invariant. It stems from the
fact that the spectrum is encircling a finite region as the wavevector $k$
varies\cite{PhysRevB.103.L140201,li2021quantized}, which introduces
directional phase accumulation in the complex energy spectrum. The offset
magnitude can be controlled by tuning the source frequency. Extending
this shift to 2-dimenional (2D) lattices, this mechanism generates
frequency-dependent V-shaped propagation patterns, where the apex
angle is dictated by the interplay between the injected light's frequency and
the strength of the interchain's coupling.

\section{Center-source offset in the 1D non-Hermitian model.}

We study a 1D microcavity lattice consisting of two sublattices, labeled
by $\alpha$ and $\beta$\cite{heebner2008optical,peng2014parity,zhang2020synthetic}. The equation of motion is given by:  
\begin{equation}
{
\begin{gathered}
i \frac{d}{dt}\binom{\alpha_n(t)}{\beta_n(t)} = 
\left(\begin{array}{cc}
iR & R \\
R & -iR
\end{array}\right) \binom{\alpha_{n-1}(t)}{\beta_{n-1}(t)} 
+ \left(\begin{array}{cc}
i\Gamma_1 & V \\
V & i\Gamma_2
\end{array}\right) \binom{\alpha_n(t)}{\beta_n(t)} \\
+ \left(\begin{array}{cc}
-iR & R \\
R & iR
\end{array}\right) \binom{\alpha_{n+1}(t)}{\beta_{n+1}(t)} 
+ \binom{if(t)\delta_{nm}}{if(t)\delta_{nm}},
\end{gathered}
}
\label{1d_equation_of_motion}
\end{equation}
where $n$ indexes the unit cell, $R$ and $V$ denote the intercell and intracell
couplings, respectively, and $\Gamma_{1(2)}$ represent the damping rates. The
final term corresponds to an external point-like source located in the $m$-th
unit cell, where $f(t)$ describes the temporal profile of the source. For
simplicity, we assume $m=N/2$, positioning it at the system's center, and set
the source frequency to $\omega_0$, such that $f(t) = e^{-i\omega_0 t}$. The
imaginary unit $i$ on both sides of the equation ensures a Schrödinger-like
structure.

\begin{figure}[ht]
	\centering
	\includegraphics[width=0.45\textwidth]{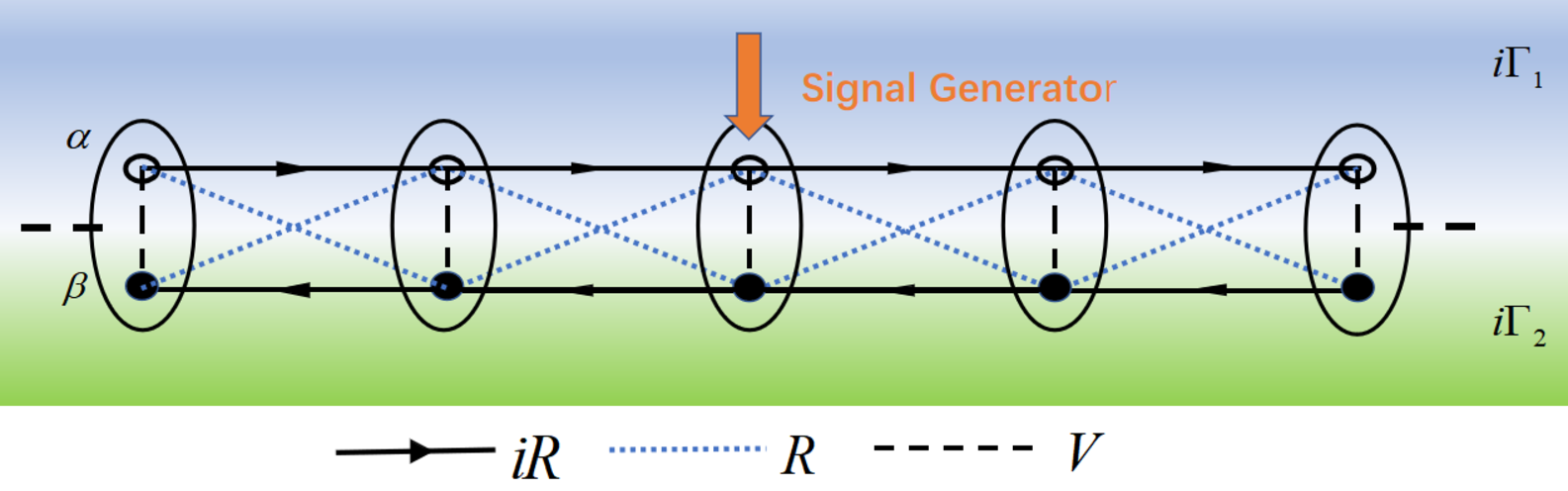}
	\caption{Schematic of the 1D non-Hermitian micro-ring resonator lattice with nearest-neighbor couplings,The arrows denote the phase direction.}
\label{fig:1dmodel}
\end{figure}

A schematic of the model is shown in Fig.~\ref{fig:1dmodel}. After applying a Fourier transform, the equation of motion becomes:  
\begin{equation}
{
\begin{gathered}
i \frac{d}{d t}\binom{\alpha_{k}(t)}{\beta_{k}(t)} = \left(\begin{array}{cc}
2 R \sin k+i \Gamma_{1} & V+2 R \cos k \\
V+2 R \cos k & -2 R \sin k+i \Gamma_{2}
\end{array}\right)\binom{\alpha_{k}(t)}{\beta_{k}(t)} \\
   +\frac{i e^{-i m k}}{\sqrt{N}}\binom{f(t)}{f(t)}.
\end{gathered}
}
\label{eq:1d_equation_after_FT}
\end{equation}

\begin{figure}[ht]
	\centering
	\includegraphics[width=0.4\textwidth]{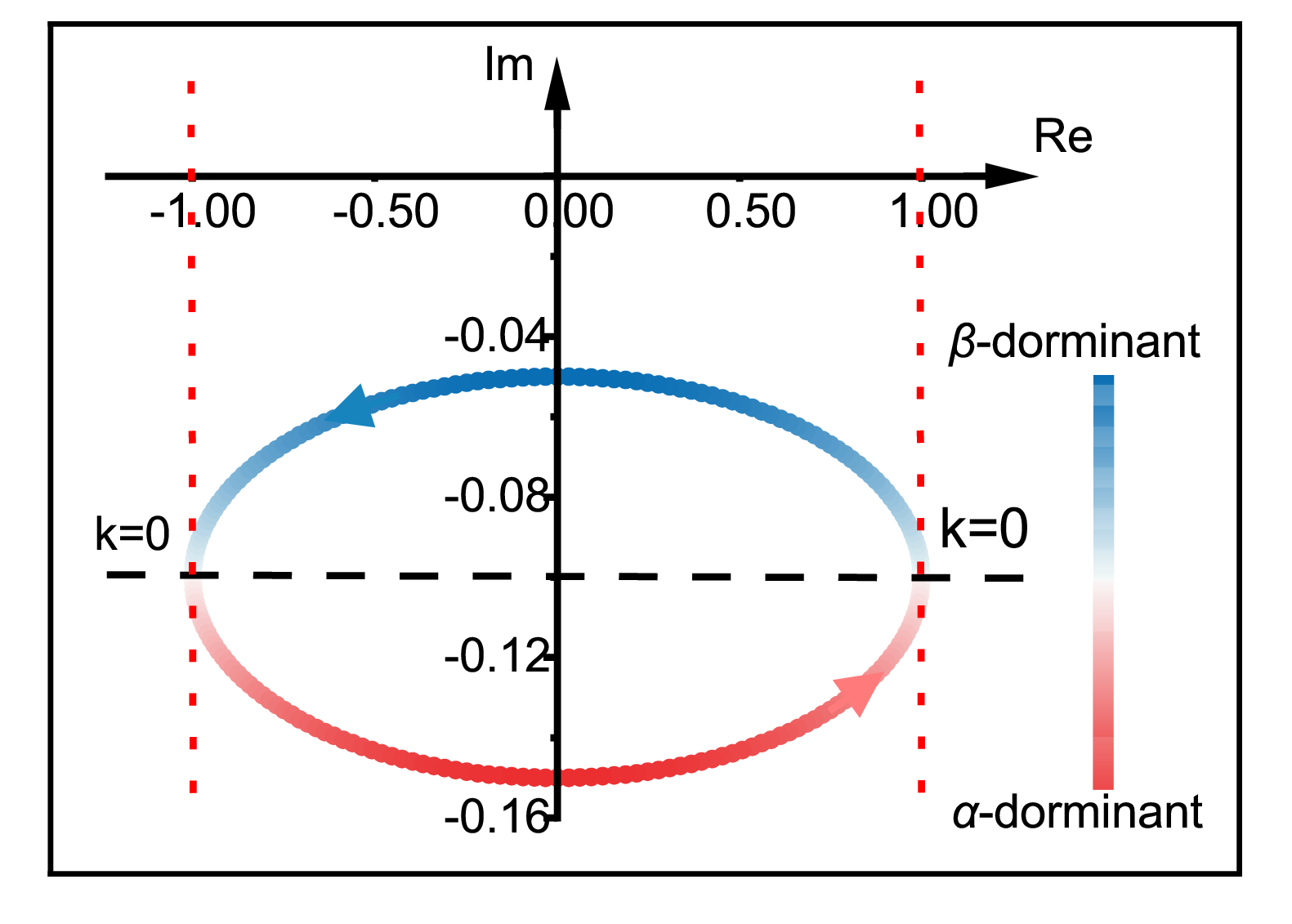}
	\caption{Eigenfrequency spectrum for $\Gamma_1 = -0.05$, $\Gamma_2 = -0.15$, $V = 0.5$, and $R = 0.25$. Arrows indicate the frequency change direction with $k$.}
\label{fig:1d_eigenfrequency_spectrum}
\end{figure}

In the absence of an external source, $H_k$, which is the $2 \times 2$ matrix in
Eq.~\eqref{eq:1d_equation_after_FT}, determines the eigenfrequencies and
eigenstates. Depending on the parameters, the eigenfrequencies can either form
two distinct bands or merge into a single unified band in the complex plane. The
findings of this paper apply to both scenarios, and for simplicity, we focus on
the latter case. As shown in Fig.~\ref{fig:1d_eigenfrequency_spectrum}, all
eigenfrequencies lie below the real axis, indicating their decay over time
and ensuring system stability. A color scheme distinguishes the
dominant sublattice in the eigenstates, with red representing the $\alpha$
sublattice and blue representing the $\beta$ sublattice.

\begin{figure}[ht]
    \centering
    \includegraphics[width=0.5\textwidth]{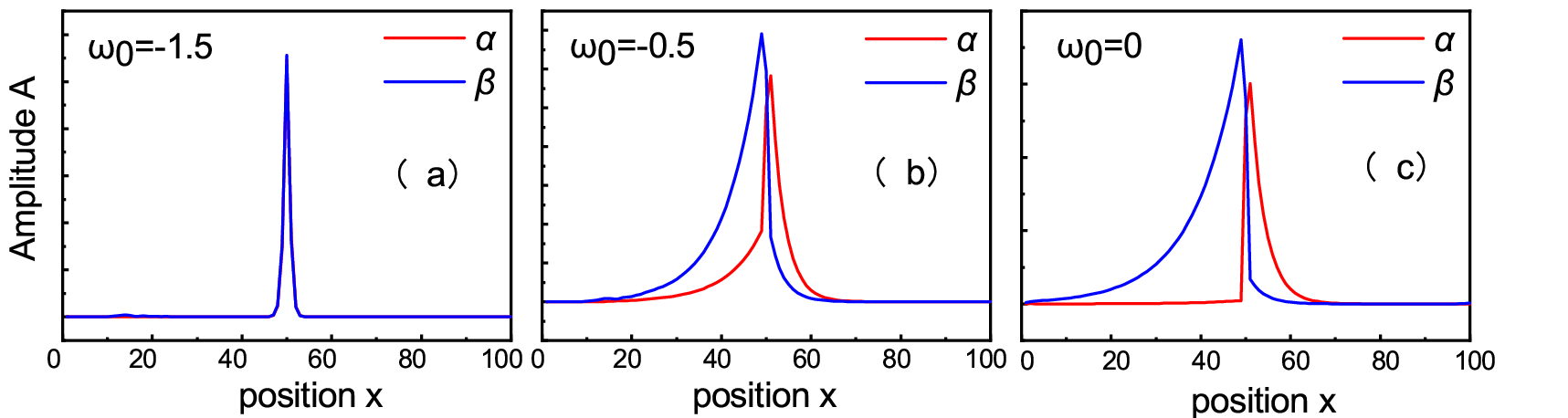}
    \caption{Position-amplitude diagrams for different $\omega_0$ at $t = 75$. Red and blue represent the $\alpha$ and $\beta$ sublattices, respectively. The lattice parameters are the same as in Fig.~\ref{fig:1d_eigenfrequency_spectrum}.}
\label{fig:amplitude_diagrams}
\end{figure}

We numerically solve the equation of motion, Eq.~\eqref{1d_equation_of_motion},
using the fourth-order Runge-Kutta method on a chain of 100 units with periodic
boundary conditions. The amplitude of the final state, which ceases to evolve
over time, is depicted in Fig.~\ref{fig:amplitude_diagrams}. At $\omega_0 =
-1.5$, the highest amplitude coincides with the source location (the 50th unit
cell). However, for $\omega_0 = 0$ and $-0.5$, the peak amplitude shifts: on the
$\alpha$ sublattice, it moves to the 51st unit cell, and on the $\beta$
sublattice, it shifts to the 49th unit cell. This shift persists when
$|\omega_0|< \Omega_0$, where $\Omega_0 =1$ defines the range of the
eigenspectrum along the real axis. Additionally, we observe that the shape of
the wave packet exhibits a bias correlated with this shift.

To clarify the observed offset, we analyze the system’s steady-state solution $\vec{\psi}$, which corresponds to the stationary solution of
Eq.~\eqref{eq:1d_equation_after_FT} after sufficient evolution time. 
To resolve ambiguities between momentum (k) and real-space (r) representations
we employ Dirac's bra-ket notation $|k\rangle$ and $\langle \langle k|$ to
denote the right and left eigenstates of $H_k$,  
The observed state is given by the real space projection $\vec{\psi}= \langle r|\psi\rangle$ of $|\psi \rangle$, which is
\begin{equation}
|\psi\rangle  = \frac{1}{\omega_0 - H} |f \rangle =\sum_{ka} \frac{1}{\omega_0 -
\epsilon_a(k)} |k_a\rangle \langle\langle k_a | f\rangle.
\end{equation}
Here $\epsilon_a$ is the $a$th eigenvalue of $H_k$, $|f\rangle$ is the external
driving term. The left and right eigenstates have the orthonormal relations:
$\langle \langle k_a | k'_{a'} \rangle=\delta(k-k')\delta(a-a')$.

We show that the state $|\psi \rangle$ is expressed as a dual-space summation
over both momentum $k$ and band-index $a$, $|\psi \rangle =\sum_{ka} Y_a (k)
\hat{Z}_a(k) |f \rangle$, where $Y_a(k)=\frac{1}{\omega_0-\epsilon_a(k)}$ and
$\hat{Z}_a= |k_a\rangle \langle\langle k_a |$. The summation over $k$
corresponds to a real-space convolution after Fourier transform, as dictated by
the convolution theorem. Notably, this formalism can be extended to the
band-index space $a$ by introducing an additional Fourier transform across
bands, thereby converting the summation over $a$ into another convolution.  This
approach is further motivated by the interconnected band structure of our model.
The energy dispersion $\epsilon_a(k)$ and the eigenstates exhibit a periodicity
of $4\pi$, effectively doubling the Brillouin zone compared to the original
$2\pi$ periodicity of $H_k$.

We introduce the $q$ specified Wannier functions $|n_q\rangle$ and $|n_q \rangle
\rangle$ as:
\begin{equation}
|n_q\rangle (|n_q \rangle\rangle) = \sum_{ka} e^{ikn} e^{iq(a-1)} |k_a\rangle (|k_a
\rangle\rangle),
\end{equation}
where $k\in (0,2\pi)$ and $q$ takes $D$ discrete numbers in $(0,2\pi)$ and $D$
is the number of bands. This procedure implements sequential band
alignment in the $k$ space, ensuring the composite spectrum spans a total
interval of $2\pi D$ without overlap. The sequential alignment optimizes
the arrangement of individual bands to collectively occupy the extended $2\pi D$
range, preserving spectral continuity while avoiding aliasing artifacts. 

The observed state $|\psi\rangle$ is
\begin{equation}
   |\psi \rangle = \sum_{nq} Y_{-q}(-n) \hat{Z}_{q}(n) |f \rangle,
\end{equation}
where $\hat{Z}_{q}(n)=\sum_{n'q'}|(n'+n)_{q'+q}\rangle \langle \langle
{n'}_{q'}|$. Because $\hat{Z}_{q}(n) \hat{Z}_{-q}(-n)=\hat{1}$, the offset matrix
$\hat{X}_{q}(n)$ which is defined by $\sum_k e^{ik\hat{X}_{q}(n)}=\hat{Z}_q(n)$
will change sign as $q\to -q$ and $n\to -n$ simultaneously. The universality of
this property is preserved irrespective of the Hermitian nature of the
Hamiltonian $H_k$. 

Here we present the analysis of three demo cases for $\epsilon_a(k)$. The first one
is that a hermitian Hamiltonian, where $\epsilon_a(k)$ resides strictly on
the real axis. This ensures that $Y_a(k)$ is a real-valued function in both $k$ and $a$
space. This reality condition enforces the symmetry: $Y_q(n)$ is an even
function in both Fourier-transformed $q$ and $n$ spaces.
Consequently, such systems inherently suppress spectral offsets due to the
absence of complex interference or asymmetric energy contributions.  The
second case involves $\epsilon_a(k)$ encircling $\omega_0$ in the complex
plane, satisfying $\frac{1}{\omega_0-\epsilon_a(k)}=e^{ik}$. After Fourier transformation,
$Y_q(n)=\delta(n+1)\delta(q)$. This indicates that all components of
$|\psi\rangle$ offset one unit cell to the right relative to $|f\rangle$. 
However, this lattice becomes dynamically
unstable due to the emergence of exponentially growing eigenstates.
The third case aligns with the model studied in this paper.
At first glance, the absence of spectral winding (i.e., $\epsilon(k)$ does not
encircle $\omega_0$) suggests no offset should occur. However, when
approximating the spectrum as a circle of radius $R_0$ centered at $O$ in the
complex plane, the terms $Y_1(k)=\frac{1}{\omega_0 - O-R_0e^{ik/2}}$ and
$Y_2(k)=\frac{1}{\omega_0 -O+R_0e^{ik/2}}$ introduce asymmetry in $k$
space. This asymmetry propagates to $Y_q(n)$ via Fourier transformation,
breaking symmetry in $q$ and $n$ spaces and enabling the offset. The strength of
the asymmetry diminishes as $|\omega_0-O|$ increases. Consequently, when
$\omega_0 =-1.5$ (far from $O$), no offset is observed,
despite the residual non-Hermitian nature of $H_k$. 

The offset phenomenon, while not a strict topological invariant due to its
dependence on the detailed spectral geometry (e.g., the components of $Y_q(n)$),
is strongly tied to how rapidly spectrum encircles the point gap.  A higher
encircling rate enhances the phase gradient in $Y_a(k)$, dominating higher-order
terms in $Y_q(n)$. For instance, doubling the encircling rate in a model with
next-nearest-neighboring hoppings leads to a doubled offset.

The offset can also be intuitively understood by analyzing the
sublattice-resolved spectral evolution in Fig.
\ref{fig:1d_eigenfrequency_spectrum}. For the first band, the spectrum rotates
anticlockwise around $\omega_0$ roughly, while the second band rotates
clockwise. Crucially, the two bands exhibit sublattice imbalance: the first band
is dominated by the $\alpha$ sublattice component, and the second band by the
$\beta$ sublattice. When the response state $\psi$ is projected onto each
sublattice separately, these opposite winding directions induce oppositely
directed offsets for $\alpha$ and $\beta$.

\section{In the Two-Dimensional Non-Hermitian Lattices.}

The above research has shown that, under specific conditions in a 1D chain, the
center of the response wave packet produced by external driving may not coincide
with the source position. This intriguing behavior prompts the inquiry: What
will be the system's dynamics when multiple such chains are arranged in parallel
and weak couplings are introduced between the adjacent chains?

\begin{figure}[ht]
	\centering
	\includegraphics[width=0.45\textwidth]{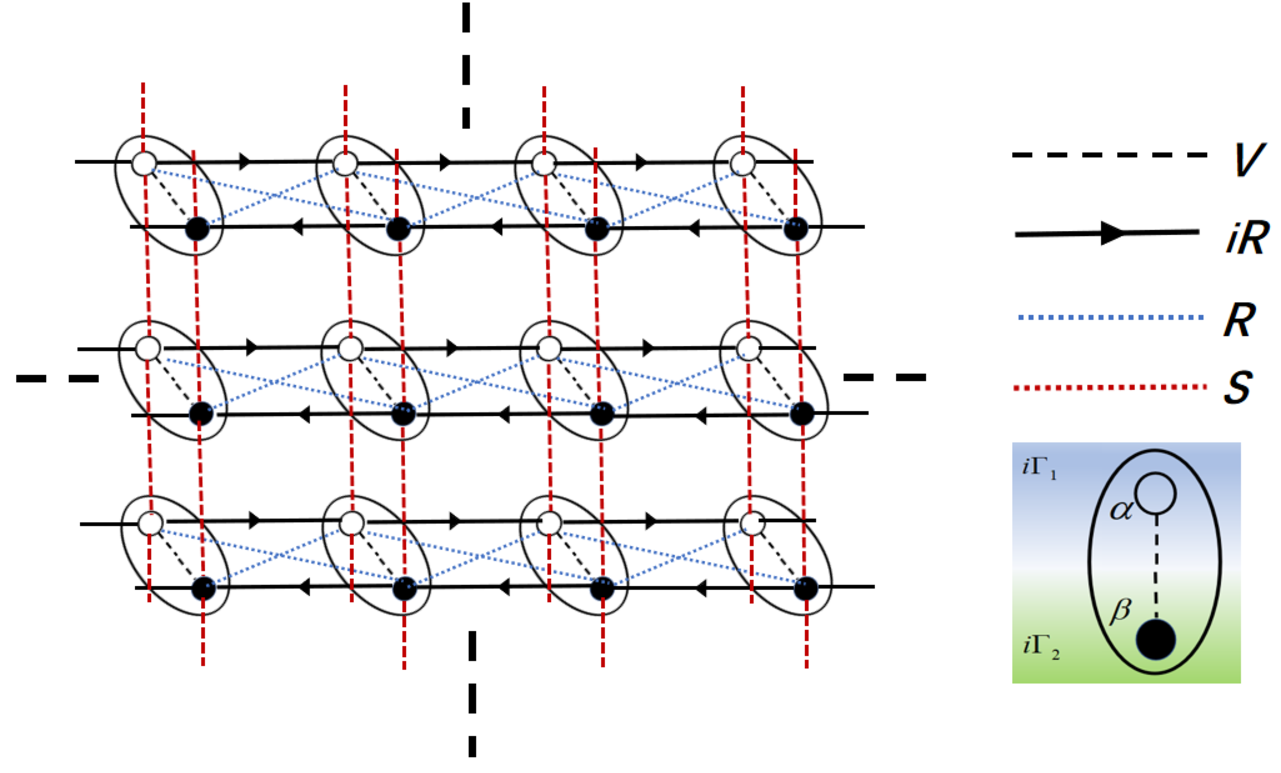}
	\caption{Two-dimensional non-Hermitian composite micro-ring resonator array with nearest-neighbor couplings.
}
\label{fig:2d model}
\end{figure}

We establish a two-dimensional non-Hermitian square micro-ring resonator array, as depicted in Fig.~\ref{fig:2d model}.
The equation of motion reads

\begin{figure}[ht]
	\centering
	\includegraphics[width=0.5\textwidth]{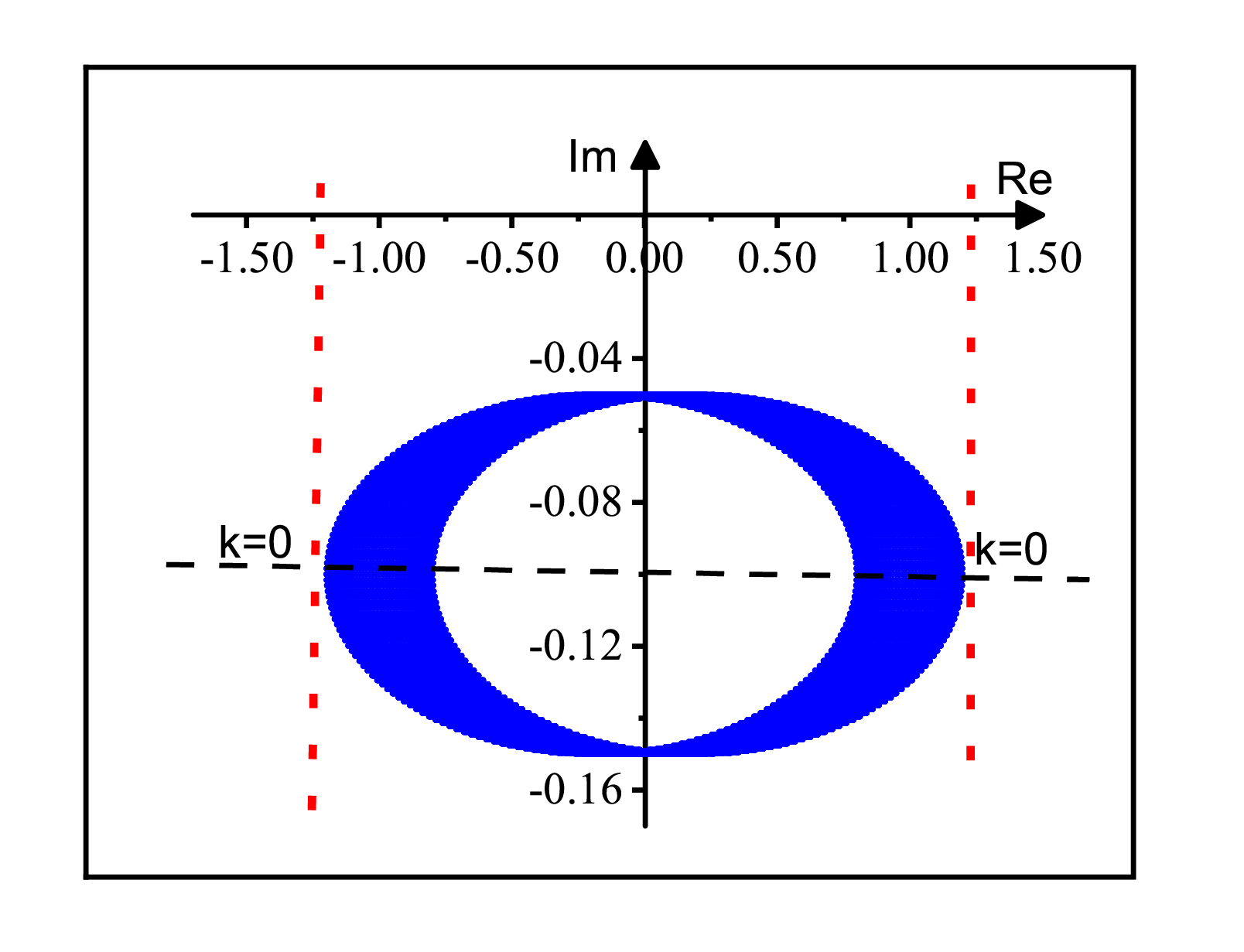}
	\caption{The eigenfrequency spectrum for $\Gamma_{1}=-0.05$, $\Gamma_{2}= -0.15$, $V=0.5$, $R=0.25$ and $\omega_0$=-0.5.
}
\label{fig:2D eigenfrequency spectrum}
\end{figure}

\begin{figure*}[ht]
    \centering
    \includegraphics[width=0.9\textwidth]{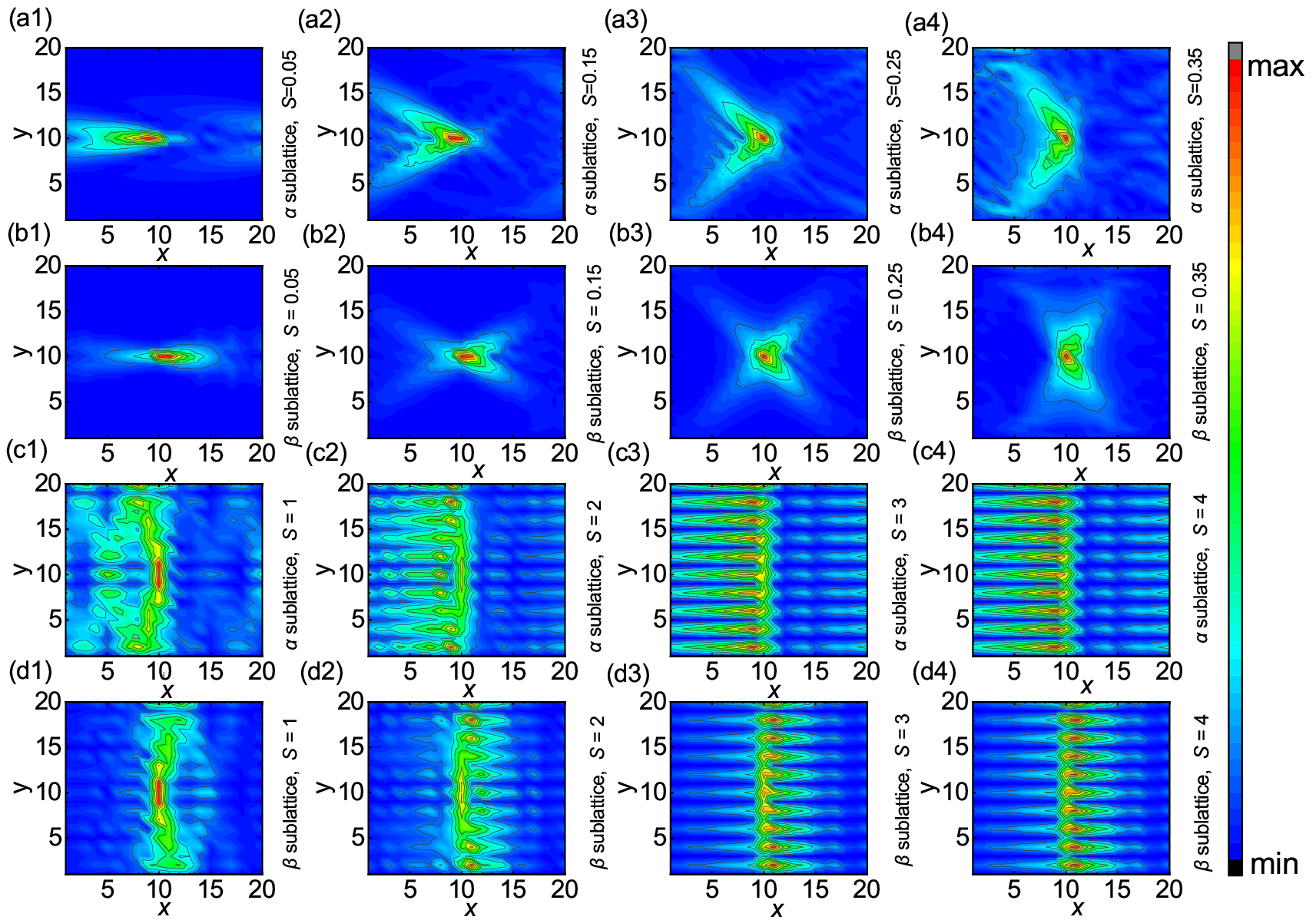}
    \caption{At $t = 50$, the amplitude of the excited wave packet on the 2D lattice. The parameters of the lattice are the same as that in Fig.~\ref{fig:1d_eigenfrequency_spectrum}. $\omega_0=-0.5$.}
    \label{fig:V-like shape}
\end{figure*}

\begin{equation}
{
\begin{gathered}
i \frac{d}{d t}\binom{\alpha_{n_x, n_y}(t)}{\beta_{n_x, n_y}(t)} =\left(\begin{array}{cc}
i \Gamma_{1} & V \\
V & i \Gamma_{2}
\end{array}\right)\binom{\alpha_{n_x, n_y}(t)}{\beta_{n_x, n_y}(t)} \\
+\left(\begin{array}{cc}
-i R & R \\
R & i R
\end{array}\right)\binom{\alpha_{n_x+1, n_y}(t)}{\beta_{n_x+1, n_y}(t)}+ \left(\begin{array}{cc}
i R & R \\
R & -i R
\end{array}\right)\binom{\alpha_{n_x-1, n_y}(t)}{\beta_{n_x-1, n_y}(t)}\\
+\left(\begin{array}{cc}
S & 0 \\
0 & S
\end{array}\right)\binom{\alpha_{n_x, n_y\pm 1}(t)}{\beta_{n_x, n_y\pm 1}(t)} 
+\binom{i f(t) \delta_{x x_{0}} \delta_{y y_{0}}}{i f(t) \delta_{x x_{0}} \delta_{y y_{0}}}
\end{gathered}
}
\label{eq:2D equation of motion}
\end{equation}
Here, $n_x$ and $n_y$ denote the indices of the unit cells on the 2D lattice. The hopping terms along the $x$-axis and the on-site damping rates are identical to those in the 1D case. The parameter $S$ represents the strength of the hopping along the $y$-axis between adjacent sites. An external source is located at $(n_{x_0}, n_{y_0})$, which is at the center of the 2D lattice in our study.

Applying a Fourier transform to the wave vector space $(k_x, k_y)$, we derive
the time-evolution equations for the system at each $(k_x, k_y)$, 
$
{
i \frac{d}{d t}\binom{\alpha_{k_{x}, k_{y}}(t)}{\beta_{k_{x}, k_{y}}(t)}=
H_{k_{x}, k_{y}}\binom{\alpha_{k_{x}, k_{y}}(t)}{\beta_{k_{x},
k_{y}}(t)}+\frac{i e^{-i x_{0} k_{x}} e^{-i y_{0} k_{y}}}{\sqrt{N_{x}
N_{y}}}\binom{f(t)}{f(t)}.
}
$
This process yields the dynamical matrix $H_{k_{x}, k_{y}}$ in $k$-space, 
\begin{equation}
H = \begin{pmatrix}
U_1 & V + 2R\cos k_{x} \\
V + 2R\cos k_{x} & U_2
\end{pmatrix},
\end{equation}
where $U_{1(2)}=\pm 2R\sin k_{x} + 2S\cos k_{y} + i\Gamma_{1(2)}$.

The spectrum of the Hamiltonian $H_{k_{x}, k_{y}}$ is given by 
$
\omega_{k_{x}, k_{y}} =2 S \cos k_{y}+\frac{i (\Gamma_{1}+\Gamma_{2})}{2} \\
 \pm \sqrt{\left(2 R \sin k_{x}+\frac{i (\Gamma_{1}-\Gamma_{2})}{2}\right)^{2} +
\left(V+2 R \cos k_{x}\right)^{2}},
$
and is illustrated schematically in Fig.~\ref{fig:2D eigenfrequency spectrum}. The system remains stable, as all spectral components lie below the real axis.

To solve the system comprising $20\times20$ unit cells subject to periodic boundary conditions, the fourth-order Runge-Kutta numerical method is utilized. This approach yields numerical solutions for the complex amplitudes at each micro-ring cavity at every time step, thereby ensuring precise tracking of the system's temporal evolution.

We observe that the $\alpha$ ($\beta$) component of the wave packet spreads
asymmetrically from the center, creating a distinct V-shaped pattern as
illustrated in Fig.~\ref{fig:V-like shape}. Notably, the angle of this V is
modulated by the magnitude of the coupling parameter $S$.

The anomalous transport observed in the 2D lattice can be attributed to the
positional offset mechanism inherent in individual 1D chains. When the
interchain coupling strength $S$ is sufficiently weak, the 2D lattice can be
approximated as a superposition of weakly coupled 1D chains. By localizing a
delta-function external source with driving frequency $\omega_0$ at the
system center, the dynamics of the central chain directly inherits the 1D
behavior. For $|\omega_0|< \Omega_0$ (where $\Omega_0$ marks
the critical frequency threshold), the excited wavepacket in the
$\alpha$-sublattice undergoes a left positional shift of one unit cell along the
chain direction, while the $\beta$-sublattice exhibits an opposite offset with less
pronounced strength.  Adjacent chains become activated through nearest-neighbor
coupling to the central chain. The external excitation in these
secondary chains comes from the excited state on the central chain and
inherits the shifted position. They will
subsequently induce an additional positional offset. 
This cascading displacement propagates radially outward, generating a
characteristic V-shaped wavefront trajectory originating from the driven
central site.  The enhanced anomalous transport in the $\alpha$-sublattice
compared to the $\beta$-sublattice arises from fundamental differences in
$\frac{1}{\omega_0 - \epsilon_a(k)}$ for the two bands. The peak for
$\beta$-sublattice in Fig. \ref{fig:amplitude_diagrams} is lower than that of
$\alpha$-sublattice, indicating stronger damping in the sublattice.
Consequently, the $\beta$-sublattice excitation
demonstrates reduced spatial coherence, rendering its V-shaped trajectory
less distinguishable compared to the sharp propagation signature of the
$\alpha$-sublattice. When $|\omega_0|>\Omega_0$, the offset mechanism in the 1D
chains disappear, as does the V-shaped wavefront trajectory.

For strong interchain coupling, the previously described transport scenario
breaks down. A representative case at $S=4$ demonstrates that the wave
morphology undergoes a marked transformation. Specifically, the externally
driven excitation at the system center exhibits anisotropic propagation: it
propagates preferentially along the vertical axis while retaining the
characteristic behavior of the 1D chain in the horizontal direction. This
phenomenon arises from the dominance of interchain hybridization over intrachain
dispersion when $S$ exceeds the critical coupling strength.

\section{Conclusions.} 
This study establishes a micro-resonator arrays to
present non-equilibrium steady-state dynamics in non-Hermitian systems under
coherent excitation, where results unveil transport anomalies:
wavepacket displacement offset in 1D lattices and V-shaped wavefront
propagation in 2D geometries.

Previous research on transport in non-Hermitian systems has predominantly
focused on the unbalanced evolution of distinct eigenstates. However, our
findings provide fresh perspectives on response dynamics within these systems.
Notably, while the offset itself is not a topological invariant, the proposed
mechanism allows for direct estimation of the rate of spectral evolution in the
complex plane. This offers a novel approach to understanding and analyzing
transport phenomena in non-Hermitian systems.

In practical applications, our work provides a methodology for elucidating the
spectral structure on the complex plane. Moreover, the fundamental mechanism we
uncover can be applied to obfuscate the source  location, enabling applications
in diverse domains such as signal processing and wavefront-engineering. For
example, these lattice configurations enable the separation of the
dual modes in an injected signal at specific frequencies.

\begin{acknowledgments}
This work was supported by the National Foundation
of Natural Science in China Grant No. 10704040.
\end{acknowledgments}



%
\end{document}